\documentclass[preprint,journal]{vgtc}       

\ifpdf
  \pdfoutput=1\relax                   
  \usepackage{graphicx}                
  \DeclareGraphicsExtensions{.pdf,.png,.jpg,.jpeg} 
\else
  \usepackage{graphicx}                
  \DeclareGraphicsExtensions{.eps}     
\fi%

\graphicspath{{figures/}{pictures/}{images/}{./}} 

\usepackage{microtype}                 
\PassOptionsToPackage{warn}{textcomp}  
\usepackage{textcomp}                  
\usepackage{mathptmx}                  
\usepackage{times}                     
\usepackage{cite}                      
\usepackage{tabu}                      
\usepackage{booktabs}                  

\usepackage[svgnames]{xcolor} 
\usepackage{soul} 
\usepackage{listings}
\usepackage{tabu}

\renewcommand{\st}[1]{} 

\definecolor{ctlchartorange}{RGB}{239,138,98}
\definecolor{ctlchartblue}{RGB}{103,169,207}
\definecolor{ctlchartgray}{RGB}{160,160,160}
\definecolor{buttongray}{RGB}{140,140,140}

\usepackage{wrapfig} 

\newcommand{\inlinebutton}[1]{\raisebox{0pt}[0pt][0pt]{\raisebox{-0.6ex}{\includegraphics[height=2.4ex]{#1}}}}

\newcommand{\paragraphheader}[1]{\noindent{\emph{#1}}~}

\newcommand{\subsubsectionheader}[1]{\vspace{1.5mm}{\noindent\textbf{#1}}~}

\newcommand{\redout}[1]{\textcolor{red}{\st{#1}}}
\newcommand{\rev}[1]{#1}

\onlineid{1530}

\vgtccategory{Research}
\vgtcpapertype{Application}

\title{MolSieve: A Progressive Visual Analytics System for\\ Molecular Dynamics Simulations}

\author{Rostyslav Hnatyshyn,
Jieqiong Zhao,
Danny Perez,
James Ahrens,
Ross Maciejewski}
\authorfooter{
\item
R. Hnatyshyn, J. Zhao and R. Maciejewski are with Arizona State University. E-mail: \{rhnatysh,jzhao,rmacieje\}@asu.edu.
\item
D. Perez and J. Ahrens are with Los Alamos National Laboratory. E-mail: \{ahrens,danny\_perez\}@lanl.gov
}

\shortauthortitle{}

\abstract{Molecular Dynamics (MD) simulations are ubiquitous in cutting-edge physio-chemical research. 
They provide critical insights into how a physical system evolves over time given a model of interatomic interactions.
Understanding a system's evolution is key to selecting the best candidates for new drugs, materials for manufacturing, and countless other practical applications. 
\redout{Thanks to recent advances in computing and simulation methodologies techniques~}\rev{With today's technology,} these simulations can encompass millions of unit transitions between discrete molecular structures, spanning up to several milliseconds of real time. 
Attempting to perform a brute-force analysis with data-sets of this size is not only computationally impractical, but would not \redout{yield clear insights about the nature of the data's physically-relevant features of the data.~}\rev{shed light on the physically-relevant features of the data.} 
Moreover, there is a need to analyze simulation ensembles in order to compare similar processes in differing environments.
These problems call for an approach that is analytically transparent, computationally efficient, and flexible enough to handle\redout{various types of physical systems} \rev{the variety found in materials-based research.}
In order to address these problems, we introduce MolSieve, a progressive visual analytics system that enables the comparison of multiple long-duration simulations. 
Using MolSieve, analysts are able to quickly identify and compare regions of interest within immense simulations through its combination of control charts, data-reduction techniques, and highly informative visual components.
A simple programming interface is provided which allows experts to fit MolSieve to their needs.
To demonstrate the efficacy of our approach, we present two case studies of MolSieve and report on findings from domain collaborators.
} 


\CCScatlist{ 
 \CCScat{K.6.1}{Management of Computing and Information Systems}%
{Project and People Management}{Life Cycle};
 \CCScat{K.7.m}{The Computing Profession}{Miscellaneous}{Ethics}
}

\definecolor{transition}{HTML}{016800}
\definecolor{pseudotransition}{HTML}{C21919}
\definecolor{statesimilarity}{HTML}{C400B9}
\definecolor{superstate1}{HTML}{76B7B2}
\definecolor{superstate2}{HTML}{59A14F}
 \teaser{
  \centering
  \includegraphics[width=\textwidth]{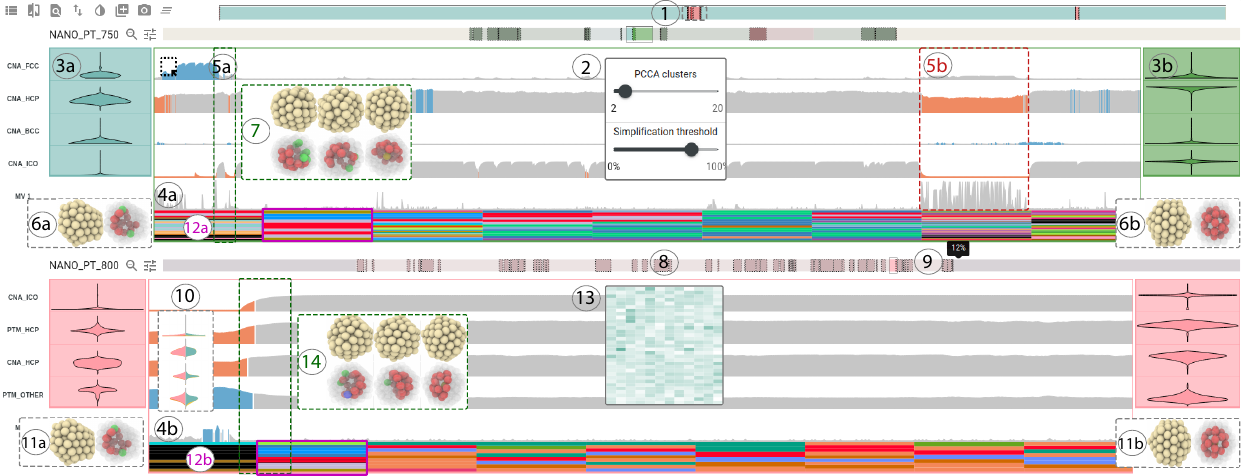}
  \caption{A sample analysis performed in MolSieve. Two long-duration simulations of nano-particles at 750 (top) and 800 kelvins (bottom) are \redout{being studied~}\rev{shown}. \rev{(}1\rev{)} \redout{displays the region that has been zoomed in on within the context of the entire trajectory in the Timeline View.~}\rev{is the Timeline View of the top trajectory before zooming.} \rev{(}2\rev{)} shows a menu for adjusting exploratory parameters of each trajectory. \rev{(}\textcolor{superstate1}{3a}\rev{)} and \rev{(}\textcolor{superstate2}{3b}\rev{)} are Super-State Views; \rev{(}6a\rev{)} and \rev{(}6b\rev{)} are their characteristic states. \rev{(}4a\rev{)} and \rev{(}4b\rev{)} \redout{show examples of a}\rev{are} multi-variate control chart\rev{s} that \redout{was~}\rev{were} dynamically added during the analysis. \rev{(}5a\rev{)} \rev{(green dashed outline)} and \rev{(}\textcolor{pseudotransition}{5b}\rev{)} \redout{show~}\rev{are} regions of interest within a Transition Region View. \rev{(}8\rev{)} shows the visible extents of the 800K trajectory. \rev{(}9\rev{)} shows the results of using the \inlinebutton{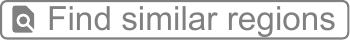} button on the transition region within the 750K simulation. \rev{(}10\rev{)} is the Region Comparison Widget for comparing the teal and pink super-states. \rev{(}\textcolor{statesimilarity}{12a}\rev{)} and \rev{(}\textcolor{statesimilarity}{12b}\rev{)} show State Space Charts\redout{,~}\rev{;} the \textcolor{statesimilarity}{boxes} display\redout{ing} similarities between the two regions. \rev{(}13\rev{)} is a Sub-Sequence Comparison Widget comparing the sequences \rev{(}\textcolor{transition}{7}\rev{)} and \rev{(}\textcolor{transition}{14}\rev{)}. An analyst has found a structural change in common between the\redout{se} simulations: \rev{(}6a\rev{)}, \rev{(}\textcolor{transition}{7}\rev{)}, and \rev{(}6b\rev{)} detail the change in the 750K simulation and \rev{(}11a\rev{)}, \rev{(}\textcolor{transition}{14}\rev{)}, and \rev{(}11b\rev{)} detail the change at the 800K simulation. \rev{(}6a\rev{)} and \rev{(}11a\rev{)} are the same state, and \rev{(}6b\rev{)} and \rev{(}11b\rev{)} are rotations of each other.}
  
  \label{teaser}
}

\begin{document}


\firstsection{Introduction}

\maketitle

Molecular dynamics (MD) simulations \redout{enable~}\rev{allow} scientists to observe how systems of atoms evolve over time using a potential energy function \redout{to~}\rev{that} calculate\rev{s} interatomic forces.
Understanding the nanoscale behavior of matter has widespread applications, from guiding protein mutations in bio-medical research~\cite{karplus1990molecular} to validating the robustness of a material in engineering contexts~\cite{massobrio2015molecular}.
A large family of software packages have been developed in order to generate MD simulations, such as GROMACS~\cite{bekker1993gromacs} for biological simulations, LAMMPS~\cite{thompson2022lammps} for materials modeling, as well as countless others, e.g.~\cite{cp2k2020, ucsfchimera2004}.\redout{With recent advances in computing capabilities, MD simulations are ballooning in size, reaching billions of timesteps in attempts to simulate processes occurring in time spans of nanoseconds.} 
A recently introduced simulation management tool called ParSplice~\cite{perez2016long} has enabled MD simulations to span \redout{even longer time-scales~}\rev{time-scales reaching into the hundreds of thousands of nano-seconds (milliseconds), two orders of magnitude larger than simulations typically performed with biological systems.
The time-scales ParSplice is able to simulate typically contain millions of discrete transitions between molecular configurations.}
\rev{Some of these systems suffer from the heterogeneous energy barrier problem~\cite{perez2016long}, a prevalent issue in long MD trajectories~\cite{Perez2015ParallelReplicaDynamics,Perez2009Accelerated}.}

Trajectories with a heterogeneous energy barrier distribution are difficult to analyze\redout{,} since relevant regions within a trajectory are buried amongst a myriad of repetitive transitions within \rev{so-called} super-states.
While calculating all of the energy paths between every state and visualizing them seems like a solution at first glance, not only are the computational costs involved impractical, but the results generated by this method \redout{would still be~}\rev{are} impossible to sift through manually.
To further compound the problem, ParSplice generates trajectories as ensembles because \redout{trajectories are~}\rev{MD is inherently a} stochastic process\redout{es}; attempting to generalize the behavior of a system\redout{~of matter} from \rev{an} individual simulation could lead to brittle conclusions.\redout{~As such, it is necessary to not only glean insights from one simulation, but to also compare its discoveries in the context of other similar simulations.}
These issues dictate the need to develop an analysis tool that highlights the essential components of a trajectory (i.e., its transition regions), while understating the parts of a trajectory where there is little to no change in the structure of the system (i.e., its super-states), as well as facilitating\redout{~efficient} comparisons between trajectories. 

\redout{There have been a~}\rev{A} number of visual analytics systems enable the exploration of molecular dynamics simulations, e.g.~\cite{Ulbrich2022sMolBoxes,Dreher2014ExaViz,Byška2019AnalysisLong,Jurcik2018Caver,Martinez.2020.VPS}.
However, most existing systems focus on biological simulations, which typically do not involve the same time-scales as their inorganic counterparts, rendering them impractical for analyzing the data-sets produced by ParSplice. 
To address this gap, we \redout{have~}worked closely with domain experts to develop MolSieve, a visual analytics system that aggressively reduces molecular dynamics simulations to their essential components (super-states and transition regions) to facilitate their analysis and comparison.
To evaluate the efficacy of MolSieve, we performed two case studies with materials science experts on \redout{diverse~}data-sets from their daily workflows. 
\redout{The case studies~}\rev{They} demonstrate that our system is not only efficient in extracting insight\redout{~from immense data-sets} but is also\redout{highly} adaptable to an expert's needs. 
This work contributes:
\begin{itemize}
    \item A novel combination of coordinated multiple views consisting of temporal charts for examining long sequences by distinguishing regions of interest and uninteresting regions\redout{, with vertically-aligned control charts to display multiple analyst-defined properties for regions of interest and violin plots to display aggregated properties for uninteresting regions, and an interaction that bridges these two types of regions by allowing experts to adjust the ir size of a region of interest};
    \item A novel state space chart for visualizing discrete temporal events in a limited screen space while outlining their general trend\redout{~which works by splitting temporal sequences and highlighting their most frequent elements};
    \item An efficient, scalable, and customizable progressive visual analytics system that supports analyzing large~\rev{materials} MD trajectory ensembles in real-time with the aforementioned visual designs.
\end{itemize}

\section{Related Work}
In this section, we review various methods to analyze long-duration molecular dynamics simulations.
We also discuss \rev{the} visualization techniques \rev{and analytical methods} that inspired our system\rev{.}\redout{~as well as the analytical methods that guided our solution.}

\subsection{Molecular Dynamics Analysis Approaches}\label{subsec:molecular_dynamics_analysis_approaches}
Many approaches exist for exploring long-duration molecular dynamics trajectories \redout{~, utilizing~}\rev{which utilize} various methods of reducing the data-set to a size tractable for real-time analysis.
We found that these approaches are typically tailored for specific analyses of biological systems. 
For example, PyContact~\cite{Scheurer2018PyContact} enables the exploration of non-covalent interactions within molecular dynamics trajectories. 
It aims to provide\redout{~quick} access to points of interest within the trajectory by filtering on the amount of contact molecules within the simulation\redout{~experience} at any given time-step.
However, PyContact requires the calculation of every\redout{~single} molecular contact before the data-set can be analyzed, which can be\redout{~computationally costly} \rev{time-consuming}.
\rev{VIA-MD~\cite{Skånberg2018VIAMD} allows the exploration of long duration biological molecular systems through a combination of linked 2D and 3D views, which work together to highlight events of interest in both the spatial and temporal domains. 
Our proposed solution differs in locating regions of interest due to the difference in scale -- VIA-MD was tested on a biological simulation that spanned twenty-three nano-seconds, while our case studies average five thousand nano-seconds. 
To extract insights from data-sets of this size, we developed a unique data simplification scheme based on the internal dynamics of the simulation. 
To the best of our knowledge, this simplification scheme has not yet been explored.}
ExaViz~\cite{Dreher2014ExaViz} enables the in-situ analysis of biological molecular systems.
This in-situ approach reduces the data-set by allowing experts to decide what portions of the trajectory are relevant before saving them for long-term storage, which requires a tremendous amount of computing power and tedious manual analysis.
Byška et al.~\cite{Byška2019AnalysisLong} built a focus+context visual analytics system that tied statistical properties of simulations to their 3D renders. 
Building on this work, sMolBoxes~\cite{Ulbrich2022sMolBoxes} utilized a data-flow model embedded in CAVER~\cite{Jurcik2018Caver} to identify important snapshots within long duration bio-molecular simulations. 
sMolBoxes identifies important snapshots (states) within a trajectory by relying on domain specific information provided by analysts, e.g., using the root-mean-square deviation (RMSD) between states to identify abnormal structural changes in proteins. 
Analysts are able to select individual parts of a protein to track throughout the trajectory.
Unfortunately, this powerful interaction is inherently coupled with the spatial dimension of the data, which reduces its scope to biological systems. 
Duran et al.~\cite{duran2019VisualizationLargeMolecular} explore building a similar system using traditional statistical charts and linking them to a 3D visualization of the protein being studied.
Non-biological systems do not behave in the same manner as proteins, reducing the effectiveness of these approaches as a general solution to identifying regions of interest within a molecular dynamics simulation.
Chae et al.~\cite{chae2019visual} used a deep learning model to reduce the dimensionality of a molecular dynamics simulation to a 3D space for easier exploration, using multiple views to display the original data alongside the 3D embedding. 
\rev{LaSCA~\cite{Tian.2023.LVA} is a visual analytics system which identifies crystalline structures within large molecular systems in great detail; however, the system does not support analyzing these structures within the context of a MD trajectory.
Wu et al.~\cite{Wu.2022.VAD} proposed a visualization pipeline to identify point defects in nuclear materials -- as with LaSCA, this approach does not consider the trajectory as a whole.}
To the best of our knowledge, the visual analytics systems currently available do not offer an efficient\redout{, generalized} method to identify and compare analyst-defined regions of interest within MD simulations \rev{of materials}. 

\redout{Along with the aforementioned visualization systems,~}\rev{A} number of programming tool-kits also provide solutions for MD trajectories ~\cite{stukowski2010atomistic,redonasamson,Brehm2020Travis, Larsen2017ASE}.
However, these tool-kits cannot identify regions of interest within a trajectory without being integrated into a larger framework. 
Blindly applying these tool-kits to long simulations without a scheme to filter and organize their output will simply produce large bodies of data that are difficult to interpret.

\subsection{Visual Analytics Methods for Time-series Exploration}
In this section, we discuss several works that directly inspired views in MolSieve.
Tominski et al.~\cite{tominski2012stacking}\redout{~provided} \rev{developed} a multi-attribute temporal view for a spatial trajectory by stacking horizon charts representing each attribute. 
This stacked trajectory chart is then rendered on top of 3D map data to facilitate a spatio-temporal analysis of the data-set.
DQNVis~\cite{wang2019dqnvis} also took a similar approach to visualize multi-variate sequence data by stacking line charts, bar charts, and area charts on top of each other to provide a multi-dimensional view of the behavior of a machine learning model. 
Additionally, their approach provides methods to identify and compare patterns within the trajectory using segment mining and dynamic time warping. 
MolSieve does not use dynamic time warping for comparing sequences, as the structure of a system is far too complex to be modeled by dynamic time wrapping; instead, we use domain-specific methods to compare analyst-defined regions.
Our approach combines the visual elements of the aforementioned systems and uses trajectory information to generate and arrange charts based on the detected importance of a region.
SignalLens~\cite{kincaid2010signallens} uses a distorted scale where interesting parts of an electronic signal are magnified while uninteresting regions are minimized in their sequence view. 
Regardless of the level of distortion,\redout{~a sense of} context is maintained, which is essential to navigating long time-series\redout{~interactively} on a screen limited by size.
MolSieve distorts the trajectory's sequence to emphasize transition regions while minimizing super-states.
\rev{~For a comprehensive review of time series visualization techniques, we refer to Aigner et al~\cite{aigner2007visualizing}.}
\newcommand{\taskheader}[1]{\vspace{1mm}\paragraphheader{\textbf{#1}}}
\section{Analytical Tasks, Requirements, and Definitions}
\begin{figure*}[t]
    \centering
    \includegraphics[width=0.95\textwidth]{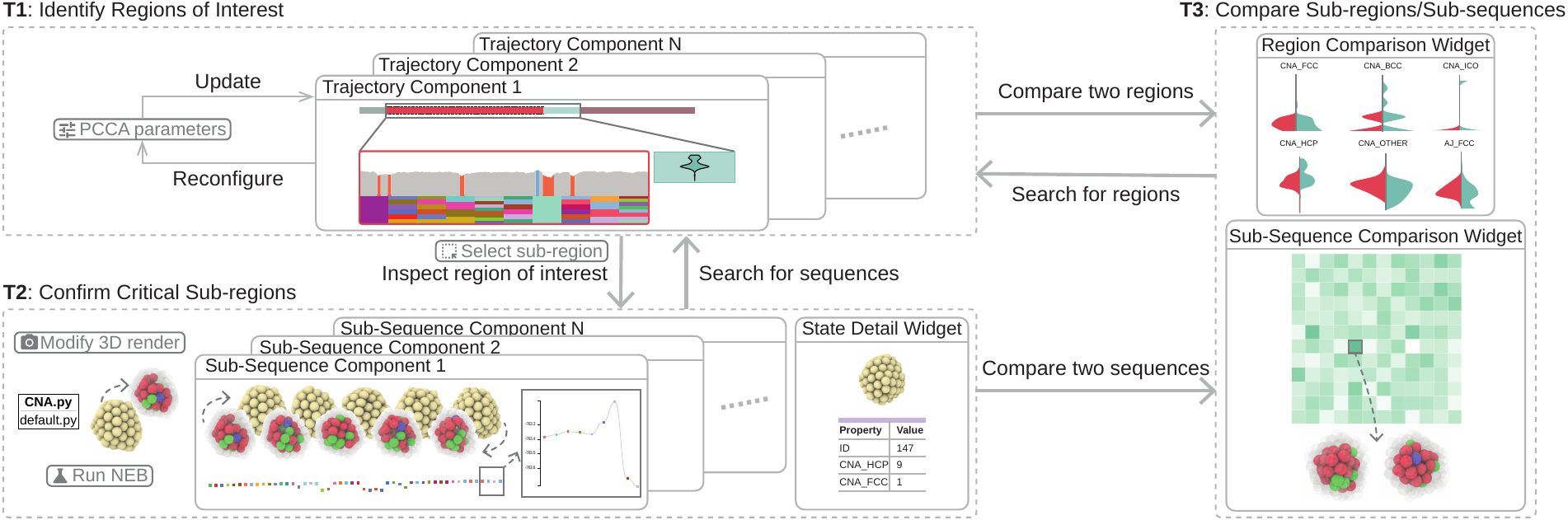}
    \vspace{-2mm}
    \caption{MolSieve is designed to extract insight from MD simulations in three stages. First, an analyst uses a modal window to set the system's exploratory parameters. When the initial simplification completes, analyst-defined properties are mined on portions of the trajectory and progressively rendered in a Trajectory Component. While the properties are being calculated, the analyst uses the Trajectory Component's embedded views to interactively identify regions of interest (\textbf{T1}). If no regions of interest are found, the exploratory parameters can be reconfigured. If the analyst finds a sub-region of interest, they can select and examine it in detail using the Sub-Sequence Component and the State Detail Widget (\textbf{T2}).
    The analyst can use the comparison interactions provided by MolSieve to explore other trajectories in the context of their new discovery (\textbf{T3}).
    \vspace{-4mm}
    }
    \label{workflow}
\end{figure*}
In this section, we define tasks for MD analysis, the requirements for an analytical tool, and domain-specific definitions.

\subsection{Definitions}\label{subsec:definitions}
ParSplice simulations typically generate tens to hundreds of thousands of unique configurations of the system being simulated, with each discrete configuration being referred to as a \emph{state}. 
Each configuration has its own state ID.
These states contain meta-data about the system being simulated at a given point in time, such as the positions, chemical species, velocities, etc.,\rev{of its atoms.}
This meta-data can be used to\redout{~gain further insight into the state by} calculat\rev{e}\redout{ing} properties that characterize its structure and geometry.

A \emph{trajectory} is a sequence of states, and a single trajectory describes one of the many possible ways a system can evolve.\redout{on a given potential energy surface.} 
To \emph{transition} between two states, the system must overcome the \emph{energy barrier} between them; therefore, state transitions that have a low energy barrier tend to occur exponentially more frequently than state transitions associated with larger barriers. 
This causes states to repeat throughout a trajectory, since structurally similar states are easier to transition to than radically different ones.
\rev{Each transition has a discrete time-step associated with it to organize it temporally; transitions take a variable amount of time, but usually they usually occur in the span of hundreds of picoseconds.}

The frequency of low energy transitions causes trajectories to often get trapped in so-called \emph{super-states}, subsets of states connected together by low energy barriers, separated from outside regions by high energy barriers. 
Parsplice simulations tend to visit these super-states for long periods of time before transitioning to another super-state. 
These movements between super-states are referred to as \emph{transition regions,}\redout{, and they} \rev{which} typically contain the most important kinetic information of a system because \rev{it controls its long-term behavior.} \redout{they control the long-time behavior of the system.} 
Transition regions are often comparatively short compared to the time spent trapped within super-states, while intra-super-state transitions occur very frequently\redout{~and typically do not control long-time evolution}.

\rev{When analyzing the structure of molecules, experts often investigate the neighbors of each atom and determine the shapes that these neighborhoods form in order to characterize a system.
Mutations in the shape and crystalline structure of a system have a strong influence on its properties.
There are seven main types of crystalline structures commonly found in materials, and our case studies are focused around analyzing cubic (face-centered cubic -- FCC, body-centered cubic -- BCC) and icosahedral (ICO) structures as they commonly occur in nano-particles; please refer to Misra~\cite{Misra.2012.C1C} for a thorough discussion.}

\subsection{Analytical Tasks}
We adopted an iterative design process to develop MolSieve\redout{~collaboratively} with two domain experts who work in computational materials science; one of them has over twenty years of experience, and the other has more than six.
We met bi-weekly for two years, using the feedback from these meetings to refine MolSieve's functionality and visual design.
Through the design process, we identified a set of analytical tasks that are essential for gaining insight into long duration molecular dynamics simulations.
Simplifying these tasks became one of the core design objectives of MolSieve \rev{(Figure~\ref{workflow})}.

\taskheader{T1: Classify super-states and transition regions in individual trajectories.}
The first step in analyzing large simulations is to identify super-states and the transition regions that separate them, which are not known \emph{a priori}.
Transition regions are critical because they control how rapidly the system will experience significant changes that could affect its properties. 
This separation reduces the data-set to a manageable size and allows experts to concentrate their analysis on transition regions.

\taskheader{T2: Identify critical sub-regions, relevant patterns and motifs within transition regions.}
\rev{There are a number of patterns and motifs to be discovered within the transition regions of a trajectory.}
Patterns of state transitions often signify the presence of a structural change, but they can also be misleading due to the nature of long duration simulations, where repeated behavior is often due to the system making rapid low-energy transitions between states. 
The \redout{primary~}challenge lies in identifying patterns and sub-regions within transition regions where meaningful changes occur while ignoring low information density portions.
The analysis of these sub-regions is the crux of molecular dynamics research; understanding how the structure of a material changes allows domain experts to make decisions on whether or not to use a certain material in an\redout{~real} engineering application.

\taskheader{T3: Compare regions of interest between trajectories.} 
MD trajectories are generated in a stochastic manner, so it is\redout{~highly} unlikely that two\redout{~different} trajectories will contain the\redout{~exact} same behavior and physical structures.
Therefore, there is a need to develop flexible\redout{~and sophisticated} methods that can differentiate robust features of the dynamics that are common to many simulations. 

\subsection{Requirements}
After identifying the primary tasks found in MD analysis, we derived the following set of requirements for a visual analytics system.

\taskheader{R1: Guide the analyst to transition regions.}
\rev{Analysts should be guided to regions that are most likely to reveal significant changes in a system's structure.}
\redout{Typical MD simulations are extraordinarily large, dictating the need to visually guide the analyst in identifying interesting regions that are most likely to reveal significant changes in a system's structure.} 

\taskheader{R2: Automatic calculation of analyst-defined properties.} 
The trajectory should be populated with automatically calculated properties that can be defined by an analyst.
\rev{Time should only be spent computing properties for regions that are potentially interesting.}\redout{An emphasis should be made to only spend time computing properties for regions that are potentially interesting and to avoid calculating properties for states that occur in super-states.}
The results should be stored in a data-base for future use.

\begin{figure*}[t]
    \centering
    \includegraphics[width=\textwidth]{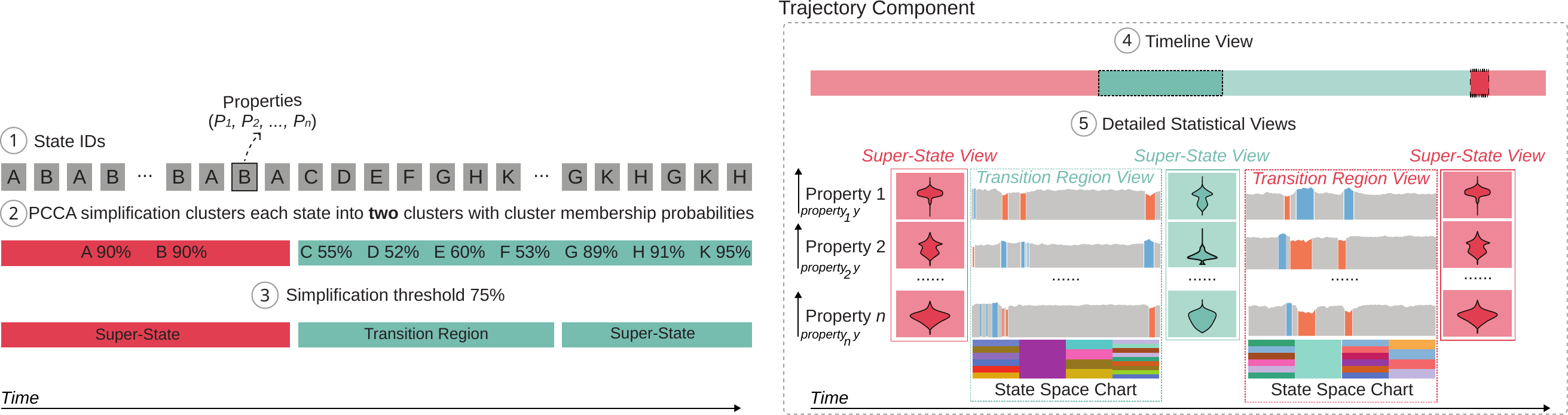}
    \vspace{-6mm}
    \caption{The simplification scheme employed by MolSieve and its relation to the visual components in the system. 
    (1) displays a portion of a sample trajectory's sequence, where each rectangle represents a state, the capital letter represents its state ID, and $P_{1}$ to $P_{n}$ represent its analyst-defined properties. 
    These properties are not required for the simplification and can be calculated and assigned afterward.
    (2) GPCCA is performed on this sequence and \rev{it} yields the maximum cluster membership probability for each state. 
    Then, the simplification \rev{(3)} is applied using an analyst-defined threshold (75\% by default). 
    States with a maximum cluster membership probability \textbf{above} this percentage are rendered as \textbf{super-states}, and states \textbf{below} are rendered as \textbf{transition regions}. These regions are mapped to views in a Trajectory Component which consists of the Timeline View and statistical views. 
    The Timeline View (4) provides temporal context for the statistical views below. Regions drawn with dashed outlines are transition regions, while regions without outlines are super-states.
    The statistical views (5) \rev{are arranged temporally and split vertically into partitions, with each partition corresponding to a single property; we include axes to indicate the relative scale for each property.}
    Super-State View\rev{s} display\redout{s a} small multiple\rev{s} of violin plots that \redout{to} outline the distributions of each property within a super-state. Transition Region View\rev{s} \redout{is a} \rev{are} small multiple\rev{s} of control charts for each property \rev{that are accompanied by~}\redout{coupled with}a State Space Chart. These charts collaborate to describe the most frequently occurring states within evenly divided segments; the number of states in a segment directly correlates \rev{to} the number of unique states visited. Segments with large numbers of states usually indicate a structural change is occurring.
}
    \vspace{-3mm}
    \label{simplification}
\end{figure*}

\taskheader{R3: Highlight potentially interesting sub-regions.}
Once the expert-defined properties are rendered, the analyst should be\redout{~visually} guided towards sub-regions within transition regions that potentially express a change in the system's behavior.
While \textbf{R2} focuses on calculating properties, guided visual exploration is another crucial aspect that accelerates the discovery process.

\taskheader{R4: Select, compare, and inspect regions of interest in detail.}
\redout{The seamless integration of~}\rev{Integrating} \textbf{R1--3} should enable the analyst to effectively select and refine regions of interest in a responsive manner, as well as allowing them to inspect a set of customized properties through expressive visualizations.

\taskheader{R5: On-demand calculation of detailed analyses.}
The selection process detailed in \textbf{R4} generates sub-regions that may include states that express behaviors of interest. Understanding their behavior requires physically grounded analyses which can \rev{be computationally expensive.}
The analyst should be able to request these analyses on demand and be able to continue exploring the trajectory.

\taskheader{R6: Extensibility.} 
An intuitive extension of \textbf{R2} is the ability to define new properties.
The solution should accommodate a broad spectrum of simulation types, enabling analysts to provide customized scripts for calculating \rev{system-specific properties.}
By providing this amount of flexibility, analysts can define properties which typically denote changes in a system.
They can \rev{then} use the visualizations and interactions provided by the solution to quickly identify regions of interest based on these properties.

\taskheader{R7: Ease of use and performance.} 
The analyst should be able to easily navigate and discern patterns\rev{within trajectories.}
Additionally, the proposed solution must remain responsive during computationally intensive tasks and progressively render partially calculated data while waiting for results. 
The analyst should receive\redout{~clear} feedback regarding the progress of complex calculations as well as any errors that may occur, with the ability to adjust or cancel \redout{complex calculations~}\rev{them} as needed.

\section{MolSieve}
MolSieve is a visual analytics system implemented using a FastAPI~\cite{FastAPI} back-end, and an interface powered by D3~\cite{D3}, React~\cite{React}, and Redux~\cite{Redux}.
The back-end provides a powerful method for simplifying \rev{dense} MD trajectories\redout{~which would otherwise remain uninterpretable}; its results are mapped to the views in the interface (Figure~\ref{teaser}).
The interface is designed to quickly guide analysts to potential regions of interest within MD trajectories (\textbf{T1}) and provides tools to interactively verify (\textbf{T2}) and compare (\textbf{T3}) multiple data-sets. 
Due to the tremendous amount of data that needs to be processed and stored on the fly, we designed our approach based on the progressive visual analytics paradigm~\cite{fekete2019progressive}. 

To support a wide range of simulations, MolSieve automatically executes, stores, and renders the results of analyst-defined Python scripts (\textbf{R2, R6}).
This feature enables analysts to specify properties that indicate a region of interest for the simulation they are studying.
These scripts are provided access to Atomic Simulation Environment (ASE)~\cite{Larsen2017ASE} representations of each state\redout{~in memory}, which can be leveraged to calculate physically relevant properties of dynamic systems, e.g., the Common Neighbor Analysis (CNA)~\cite{honeycutt1987molecular} counts for atomic structures.
These $n$ properties are calculated and assigned to each state within the trajectory (Figure~\ref{simplification}.1).
To further accelerate the process of discovery, these properties are calculated and rendered progressively, allowing analysts to gather insights throughout the data-set without having to wait for computations to finish\redout{,} (\textbf{R7}).
\redout{Analysts can also programmatically modify the way 3D views in MolSieve are rendered, which can highlight properties of interest in the molecular structure.} 

\subsubsectionheader{Background - Trajectory Simplification}\label{subsec:trajectorysimplification}
We used Generalized Perron Cluster Cluster Analysis (GPCCA)~\cite{Reuter.2018.GMS} as implemented by pyGPCCA~\cite{reuter.2022.PGP} as the basis for MolSieve's simplification scheme; GPCCA is a generalization of the robust Perron Cluster Cluster Analysis (PCCA+)~\cite{deuflhard2005robust}.
PCCA+ has been proven to accurately simplify MD trajectories by clustering together groups of kinetically linked states~\cite{Huang2017ClusterAnalysis, huang2018shapefluctuation}. 
GPCCA can be applied to simulation\rev{s}\redout{data-sets} where transitions are modeled as a Markov chain.

MolSieve simplifies the \redout{simulation's~}trajectory by dividing \redout{the entire trajectory~}\rev{it} into tentative transition regions and super-states. 
This is achieved by first running GPCCA on the trajectory, which divides it into $N$ dominant super-states, referred to as \emph{clusters}. Here, dominant super or macro-states denote meta-stable states, in the case of reversible dynamics, or, e.g., cyclic states, in the case of non-reversible dynamics~\cite{Reuter19Generalized}.
GPCCA assigns a vector of $N$ \emph{cluster membership probabilities} to each individual state which describes how strongly\redout{~the state} \rev{it} belongs to each cluster (Figure~\ref{simplification}.2).
Then, each individual state's \rev{membership} probability is compared to a threshold set by the analyst \rev{(Figure \ref{simplification}.3)}; if\redout{~a state's} \rev{its} maximum membership probability is \textbf{above} the threshold, it is considered part of a \emph{super-state}\rev{; otherwise, it is considered to be\redout{~and states \textbf{below} the threshold are considered} part of a possible \emph{transition region} (i.e., it occurs in regions where the trajectory moves between clusters).}
If the simplification threshold is set to its maximum value of 1.0, no portion of the trajectory will be simplified, and every state will be considered a transition region.

When initially loading a trajectory, analysts have the opportunity to set a range for the GPCCA clusterings they are interested in, as\redout{~they are} \rev{GPCCA is} not guaranteed to yield results for all numbers of clusters.
The back-end uses the range to determine and return the optimal GPCCA clustering for the trajectory and then simplifies it using the simplification threshold.
Simultaneously, analyst-defined properties ($P_{1}$ to $P_{n}$) are calculated and assigned to each state within the trajectory.

The optimal clustering may not always reveal the best possible splits between transition regions and super-states, so analysts are free to adjust the GPCCA cluster counts as well as the simplification threshold within the interface.
The simplification threshold is set to a default value of 0.75 and the GPCCA clustering range to 2-20 which provides a reasonable starting point for exploration.
A simplification threshold value of 0.75 tends to reveal sets of states that are weakly clustered, regardless of the GPCCA cluster count.
The default GPCCA clustering range is set wide enough to ensure a clustering is found.
Once a trajectory is simplified, its results are directly mapped to MolSieve's Trajectory Components (Figure~\ref{simplification} \rev{right}). 

\subsection{Trajectory Components}\label{subsec:trajectorypanel}
\redout{MolSieve is designed around a number of~}Trajectory Components \redout{that~}adopt a focus+context approach~\cite{furnas2006fisheye} to assist analysts in identifying regions of interest through the use of a variable number of Transition Region and Super-State Views (Figure~\ref{workflow}.T1).
Each trajectory belongs to a separate component, organized on the main area of the screen.
The Timeline View (Figure~\ref{simplification}.\rev{4}\redout{~3}) provides temporal context \rev{and a means of control} for the statistical views (Figure~\ref{simplification}.5).

\subsubsectionheader{Timeline~View}\label{subsec:timelineview}
\rev{The Timeline View (Figure~\ref{simplification}.4) displays the regions that are currently being rendered as statistical views (Figure~\ref{simplification}.\rev{5}) and allows experts to adjust which regions are visible to focus their analysis.
Regions are colored according to the GPCCA cluster they are assigned; transition regions are rendered with a dashed outline and super-states with no outline and a slightly lighter color in order to differentiate between them. 
We colored the clusters with a color scheme adapted from ColorBrewer's~\cite{Harrower2003ColorBrewerorg} qualitative set. 
Hovering over either type of statistical view highlights its corresponding region in the Timeline View.
Brushing the view adjusts the visible extent of the trajectory, saturating regions that are outside of the brush's extent and reorganizing the statistical views.
Double clicking the view zooms it in on the currently brushed region, which allows analysts to view regions that may have been rendered too small initially. 
There are two additional interactions provided by buttons next to each Timeline View.
Clicking the \inlinebutton{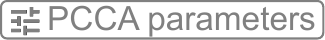} button shows a menu containing two sliders that adjust the number of GPCCA clusters and the simplification threshold (\textbf{R1}).
The \inlinebutton{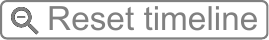} button resets the Trajectory Component to show the entire trajectory.
}

The detailed statistical views (Figure~\ref{simplification}.\rev{5}) are arranged in temporal order from left to right and are drawn with a scale that exaggerates the size of transition regions to direct the analyst's attention (\textbf{R1}).
Transition regions are exaggerated because they contain details on how the system evolved within a critical region, which demands more screen space; meanwhile, super-states are small multiples of violin plots which remain legible at small sizes.
As a result, super-states are only offered a maximum of 10\% of the total screen space unless there are no transition regions within the visible extents of the trajectory, in which case they occupy the entire width of the screen.
This exaggerated scale was inspired by SignalLens~\cite{kincaid2010signallens}.

Each view is bordered by the color of the cluster it is associated with, which assists in finding transition regions between clusters. 
The slices marked Property$_{1}$ to Property$_{n}$ in Figure~\ref{simplification}.\rev{5} display how views within Trajectory Components are vertically split into partitions.
\rev{Each property is assigned a partition per trajectory, and each property}\redout{~Each property} is consistently rendered with its own scale in order to facilitate comparison. 
Partitioning the views in this manner allows experts to follow the evolution of multiple properties simultaneously.

We implemented a dynamic ranking system for each partition, which reduces the amount of irrelevant data on the screen in order to streamline intra-trajectory comparisons.
Properties that change dynamically throughout a simulation are more likely to be relevant to analysts than properties that stay constant, so each property partition is ranked vertically based on the magnitude of its difference throughout the trajectory.
This difference is calculated by performing a statistical z-test between each pair of adjacent super-states and summing \rev{them} over the trajectory to get the final score.
Each set of rankings is individual to a trajectory, as certain properties can be highly dynamic in one trajectory and stagnant in another.
Since the data is being loaded progressively, the rank is calculated based on the information currently available to MolSieve.
By default, only 4 properties are loaded, but there is a \inlinebutton{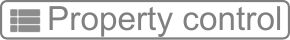} button in the main toolbar to adjust the number of properties shown.

\subsubsectionheader{Transition Region View}\label{subsec:detailedsequenceview}
To aid analysts in discovering sub-regions of interest within a transition region, we designed the Transition Region View (Figure~\ref{simplification}.\rev{5}) to leverage\redout{~the well-known ability of} control charts\redout{~to} \rev{for} detecting statistical anomalies~\cite{shewhart1939statistical}. 
Each view contains a small multiple of control charts which correspond to analyst-defined properties\redout{~working in tandem} \rev{which work} to highlight regions that have drastic changes in value.
At the bottom of each view, a chart displays the state space within the region, which provide an overview of the most frequently occurring states.\redout{~in evenly divided segments.}

\paragraphheader{\rev{Control Charts}}
Each control chart displays the moving average of\redout{one} \rev{a} property inside a transition region and it is colored based on the distance of the moving average from the mean. 
If the moving average moves one standard deviation \textbf{above} the mean, it is colored \textcolor{ctlchartblue}{\textbf{blue}}. 
If the moving average moves one standard deviation \textbf{below} the mean, it is colored \textcolor{ctlchartorange}{\textbf{orange}}, and if it stays \textbf{within} the control limits, it is colored \textcolor{ctlchartgray}{\textbf{light gray}}.
This coloring scheme, inspired by ColorBrewer~\cite{Harrower2003ColorBrewerorg}, draws attention to sequences within the transition region where a change is occurring, and allows analysts to quickly determine what sub-regions are of special interest, fulfilling \textbf{R3} and \textbf{R4}. 
Hovering over a control chart displays a tooltip with the current value of the property and associated time-step.

\begin{figure}[t]
    \includegraphics[width=\columnwidth]{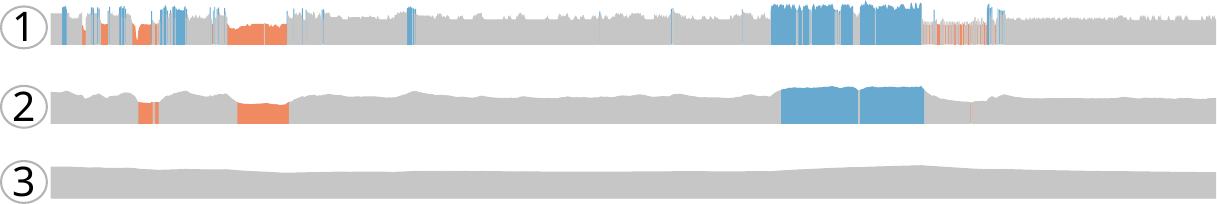}
    \caption{The importance of being able to dynamically adjust the moving average period for the Transition Region View. 
    (1) has a moving average period that is too short and contains a lot of false positives for anomalies. 
    (2) has a moving average period that is appropriately chosen for the region being studied, showing only two major anomalous events within the given time-frame. 
    (3) has a period that is too large to capture the interesting events occurring within the region.} 
\label{mva}
\end{figure}

By default, the moving average time period for each control chart is set to one tenth of the length of \rev{its} Transition Region.\redout{~it describes; this is the same length as the segments in the State Space Chart.}
However, if the analyst finds that the moving average time period is not capturing regions of interest, they can adjust the moving average time period for all of the control charts within the view (Figure~\ref{mva}).

\begin{figure}[t]
    \centering
    \includegraphics[width=\columnwidth]{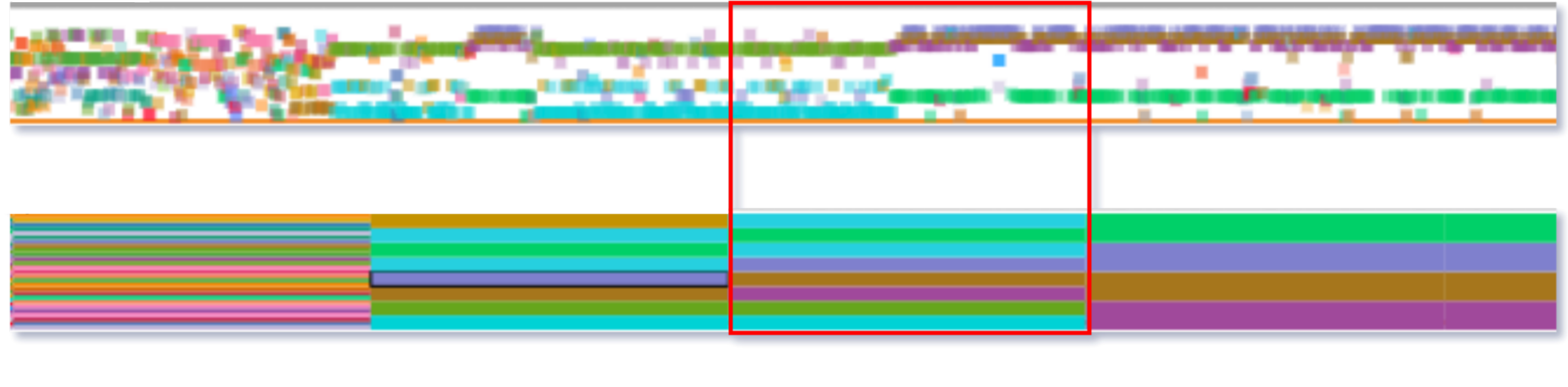}
    \vspace{-6mm}
    \caption{The original design (top) for state ID vs. time-step charts was flawed, as it attempted to render a large amount of data in a small space. 
    States would often occlude each other, and each render would be computationally costly, as each rectangle is an SVG element. 
    Our final design (bottom) underlines the behavior of a sub-region succinctly and makes it easier to read and interpret when many states are in a region. Note the highlighted segment, which captures the transition between two minor repetitive sub-regions.}
    \vspace{-4mm}
    \label{scatterplot_comparison}
\end{figure}

\paragraphheader{State Space Chart}
A state ID vs time-step pixel plot is a familiar way of visualizing ParSplice trajectories \cite{Huang2017ClusterAnalysis}.\redout{, since} \rev{E}ach time-step within a ParSplice simulation corresponds to one state. 
Rendering states this way allows analysts to quickly determine regions of interest.\redout{~as well as identify behavioral patterns inside a transition region.}
Through our iterative design process, we found that rendering each\redout{~individual} state within a transition region would lead to highly cluttered and cumbersome graphs, since transition regions often consist of thousands of states (Figure~\ref{scatterplot_comparison} top).
\rev{In order to address these issues,} we devised and implemented an aggregate version of this plot, the State Space Chart (Figure~\ref{scatterplot_comparison} bottom).

The State Space Chart\redout{~(bottom of Figure 4) is designed to} highlights changes within the transition region by splitting it into ten evenly divided segments and calculating which states occur most frequently within each segment. 
\rev{To be considered part of a segment, a state's actual distribution value needs to be greater than its expected value.
This is defined as 1 divided by the number of unique states within the segment; i.e., a state occurs equally likely as all of its neighbors.}
Segments with many colors indicate that the simulation is rapidly moving between many unique atomic configurations.
Using the control charts coupled with this view allows analysts to quickly estimate the visit frequency within the region and identify sub-sequences worthy of a detailed inspection (\textbf{T1}).

\redout{To draw a segment, the state distribution within each segment is first calculated.
To be considered part of a segment, a state's actual distribution value needs to be greater than its expected value.
This is defined as 1 divided by the number of unique states within the segment; i.e., a state occurs equally likely as all of its neighbors.
Once all of the segments have been calculated, they are rendered as a group of rectangles which correspond to the states IDs above the expected distribution value within the segment.
Aggregating items in this manner is a commonly used data-reduction technique.}

\begin{figure*}[t]
    \centering
    \includegraphics[width=\textwidth]{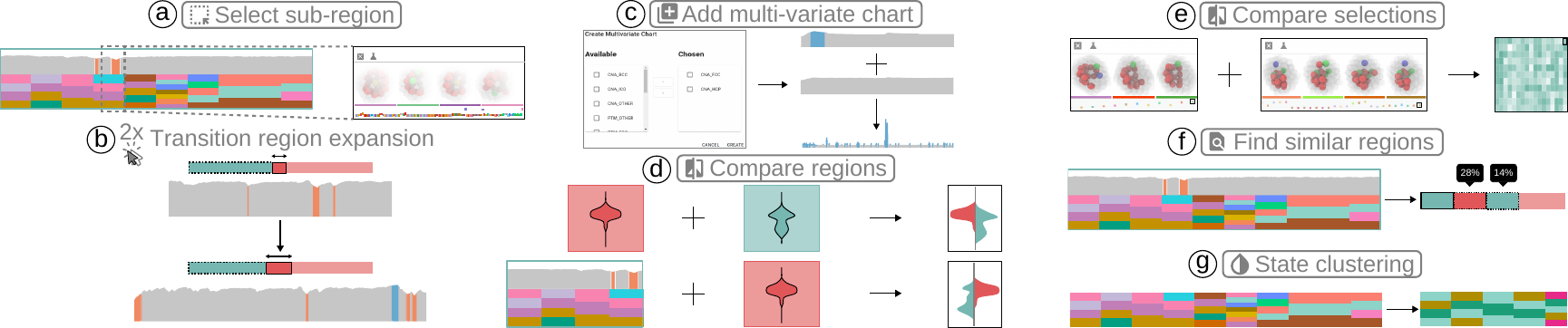}
    \vspace{-2mm}
    \caption{MolSieve's unique interactions.\rev{
    (a) lets analysts select sub-regions of interest within a Transition Region View to create a Sub-Sequence Component. 
    (b) Double clicking a Transition Region View causes it to expand into its neighbors, which makes it possible to view parts of super-states in detail. 
    (c) allows experts to create multi-variate charts. The \inlinebutton{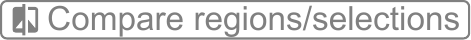} button can compare any Transition Region or Super-State View by creating a Region Comparison Widget (d) which is a small multiple of asymmetrical violin plots detailing the distributions from each selected region. 
   The button also works with Sub-Sequence Components, creating a Sub-Sequence Comparison Widget (e) that contains a heat-map detailing the similarities between the selections.
   (f) allows experts to select a single Transition Region View, which MolSieve uses to compare with all other visible transition regions, automatically highlighting their similarity on the Timeline View. 
   (g) recolors all of the states in the interface according to the OPTICS clustering algorithm using analyst-defined properties.}}
    \vspace{-4mm}
    \label{interactions}
\end{figure*}

\paragraphheader{\rev{Transition Region View Interactions}}
Upon hovering over a Transition Region View, a toolbar with multiple controls is displayed. The \inlinebutton{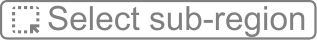} button toggles a brush to select sub-regions of interest\rev{; completing the selection generates a corresponding Sub-Sequence component (Section ~\ref{subsec:sub-sequence-panel}}; \textbf{R4}; Figure~\ref{interactions}\rev{a}\redout{~left})
To facilitate making fine-grained selections within a region, the \inlinebutton{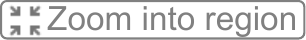} button enlarges the transition region so it occupies the entirety of the screen\redout{'s width}. 
\redout{Clicking a state in the State Space Chart updates the State Detail Widget (see Section), which statically displays the state's 3D structure as well as the value of each associated analyst-defined property.}

\subsubsectionheader{Super-State~View} 
Super-states revealed by the simplification algorithm tend to \redout{be exceedingly large, often constituting}\rev{constitute} the majority of\redout{~very large} \rev{ParSplice} simulations.
To\redout{~reduce the amount of data being processed by MolSieve and} maximize performance, we elected to use aggregate statistical charts~\cite{wickham2021boxplots} when representing super-states. 
A Super-State View (Figures~\ref{teaser}.3a,~\ref{teaser}.3b,~Figure~\ref{simplification}\rev{.5}\redout{~left}) is a small multiple of violin plots that describes the overall distribution of each\redout{~analyst-defined} property. 
They\redout{~are intended to} highlight the evolution of each property throughout a simulation in a compact manner.
\rev{We originally used box-plots to display the distributions of these properties, but we found that they were highly cluttered due to the small amount of screen space allotted to them and they did not capture the variance of each distribution as well as violin plots.}

Each violin plot is constructed using the property values from the so-called \textbf{dominant} states of the region plus a randomly selected 1\%.
Dominant states within the super-state are states that occur with larger than median frequency.\redout{, i.e., the median of all of the number of occurrences of the states within the region.}
Using a small\redout{~representative} \rev{randomly sampled} portion of the region provides a reasonable overview without having to compute a prohibitive amount of data. \redout{and is a large contributing factor to the scalability of the system.} 
\rev{In order to ensure that important details about states are not hidden from analysts, we implemented a expansion feature that allows experts to explore super-states in more detail.
Double clicking any of the control charts causes the Transition Region View to ``expand" (Figure~\ref{interactions}\rev{b}\redout{~bottom right}), revealing its state space and the moving averages of its neighbors.
Expansion occurs 100 time-steps at a time to avoid loading unnecessary data.}
\redout{Clicking anywhere in the Super-State View updates the State Detail Widget with the state that occurs most frequently within the region, referred to in as the region's \emph{characteristic state}.
This interaction allows analysts to quickly determine if any structural changes occurred between super-states, prompting them to inspect a Transition Region View in detail.}

\subsubsectionheader{Interactions}
The toolbar above all Trajectory Components provides interactions that\redout{~are intended to} enhance the analyst's ability to examine a trajectory in detail (Figure~\ref{teaser}, top left).
Analysts are able to construct multi-variate control charts~\cite{lowry1995review} with the properties they provided by clicking the \inlinebutton{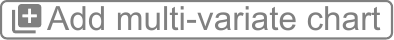} button \rev{which opens a modal window (Figure ~\ref{interactions}\rev{c}\redout{, top right})}.
These multi-variate charts (Figures~\ref{teaser}.4a,~\ref{teaser}.4b) are dynamically added to each Transition Region View, allowing the analyst to combine various properties to generate more powerful control charts that \rev{highlight synchronized movements across property values that are difficult to detect using single variable charts}\redout{~may reveal more than single-variable control charts would on their own}, fulfilling \textbf{R5}.
The \inlinebutton{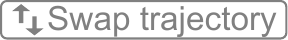} button allows analysts to swap the vertical positions of Trajectory Components to facilitate direct comparisons.
The \inlinebutton{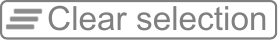} button allows experts to undo a selection they are currently making if they decide they want to abort the process.

\subsection{Sub-Sequence Component}\label{subsec:sub-sequence-panel}
Since the State Space charts within Transition Region Views only provide an overview, there is a need to look at sub-regions in more detail.
Sub-Sequence Components (Figure~\ref{workflow}.T2) are added to the bottom of the screen once an analyst completes a selection in a Transition Region View using the \inlinebutton{figures/buttons/selection_button_withtext.pdf} button (\rev{Figure} \ref{interactions}\rev{a}). 
They are designed to fulfill \textbf{R4} and \textbf{R5}, as they allow experts to glean additional insight from regions that they deem to be interesting, and correspond to the abstract/elaborate interaction category in Yi et al.~\cite{yi2007toward}.
Each Sub-Sequence Component provides a small multiple of 3D state visualizations, which serves as an overview of the structural changes occurring within the selection.
\redout{In order~}\rev{T}o generate the overview, we developed a greedy search\redout{~method} \rev{algorithm} that uses the \emph{Frobenius norm} (provided by ASE~\cite{Larsen2017ASE}) of the spatial distance between all atomic coordinates. 
\redout{Since this is a metric based on the atoms of the system being studied,~}\rev{A} high distance between states indicates that they are structurally different.
The\redout{~greedy search} \rev{algorithm} iterates over the selection and takes the distance between the state being queried and the rest. 
To find states that are as different as possible, we start at the initial state of the selection, find its most dissimilar counterpart, and start the search again at this state until we reach a maximum iteration count or the end of the selection.

At the bottom of each Sub-Sequence Component is a traditional state ID vs. time-step plot of the selection's constituent states (Figure~\ref{scatterplot_comparison} top).
The Sub-Sequence Component also supports running the Nudged Elastic Band calculation \cite{jonsson1998nudged} using the \inlinebutton{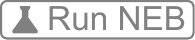} button. 
Clicking the button (Figure~\ref{workflow}.T2) opens a modal \rev{window} that allows analysts to adjust the parameters of the calculation and make a selection on the sub-sequence that will be used in the calculation.
The results from the calculation are used to generate a potential energy graph \rev{which shows} the minimum energy pathways for the selection they made, fulfilling \textbf{R5}. 
Potential energy graphs are commonly used by analysts to determine if a sequence of states constitutes a structural change in the simulation.
An exceedingly high potential energy barrier between any two pairs of states in the sequence followed by any number of low energy barriers, usually indicates a transition. 
This is because particles are known to move towards their lowest energy configurations.

\subsubsectionheader{State Detail Widget}\label{subsec:statedetailview}
Whenever a state is clicked throughout the UI \rev{(e.g., within State Space Charts, Sub-Sequence Components etc.)}, the State Detail Widget (Figure~\ref{workflow}.T2) is updated. 
It displays a \rev{static} 3D visualization of the state, inspired by guidelines outlined in By{\v s}ka et al.~\cite{Byška2019AnalysisLong} that suggest linking 3D visualizations of a system to its properties.
Additionally, a table is shown below the 3D render displaying the properties of the state that was selected.
Since states are all colored consistently throughout the visual interface, we included a bar under all 3D renders that displays the selected state's color, making it easy to visually link the state to other visualizations.
\rev{The \inlinebutton{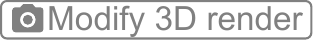} button in the trajectory toolbar allows analysts to change the way states are rendered in 3D throughout the interface by \redout{providing} Python scripts inside the \lstinline{vis_scripts} folder in the source code (\textbf{R6}\redout{; Figure.T2}). 
Experts pick the visualization script they want to use with a pop-up menu that is populated with the contents of the \lstinline{vis_scripts} folder.
Analysts are expected to define a function that takes an OVITO~\cite{stukowski2010atomistic} rendering pipeline object as a parameter which they can modify to suit their needs.
Figure~\ref{workflow}.T2 demonstrates an example: the default view is swapped for a visualization of crystalline structure neighborhoods where each atom is colored according to its structural classification (see Section~\ref{subsec:definitions}).
Customizing the visualization gives analysts an additional method to verify their conclusions made from the 2D charts in MolSieve and is integral to certain types of analyses (Section~\ref{subsec:defectanalysis}).}

\subsection{Comparison Widgets and Interactions}\label{subsec:interactions}
MD ensembles are practically impossible to analyze due to the\redout{~overwhelming} amount of data that needs to be compared.\redout{~from each trajectory.}
To address this, we included a variety of comparison interactions that quantify the difference between regions of interest from multiple trajectories (Figure~\ref{workflow}.T3 and Figure~\ref{interactions}).

The \inlinebutton{figures/buttons/compare_region_withtext.pdf} button allows experts to select regions or sub-sequences they want to compare directly\redout{~(Figure 6 middle)}. 
When two\redout{~regions} are selected, a Region Comparison Widget is placed at the bottom of the screen \rev{which contains} asymmetrical violin plots \rev{that} compare the distributions of each property \rev{(Figure~\ref{interactions}d)}.\redout{within the two regions.}
Transition Region Views can also be selected with this interaction, making it easy to compare them with Super-State Views\rev{; MolSieve uses the properties from the dominant states in transition regions to compute the distribution of each property, allowing for a fair comparison.}
Comparing regions this way reduces the cognitive load of having to look back and forth between two distributions that are visually separated.
When two Sub-Sequence Components are selected, a Sub-Sequence Comparison Widget is generated, which displays a state similarity heat-map (Figure~\ref{interactions}\rev{e}\redout{~middle bottom}). 
State similarity is defined as the inverse of the distance used in the 3D overview for Sub-Sequence Components, see Section~\ref{subsec:sub-sequence-panel}\redout{~for more details}. 

The \inlinebutton{figures/buttons/find_region_withtext.pdf} button lets an expert select a Transition Region View to quickly compare to all other \redout{visible} Transition Region Views \rev{that are currently selected using the Timeline View's brush}, which corresponds to a Connect interaction in Yi et al.~\cite{yi2007toward}.
Once the selection is complete, MolSieve computes the difference between their state distributions and then displays the result with a tooltip rendered above each region \redout{that was compared in the Timeline View.}(Figure~\ref{interactions}\rev{f}\redout{~left middle}).
This computation provides a crude preview of similarities between two transition regions, which can be used to narrow down which regions require an in-depth comparison. 

Clicking the \inlinebutton{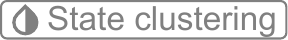} button clusters all of the states in the transition regions visible on the screen based on their properties (Figure~\ref{interactions}\rev{g}\redout{~top left}). 
MolSieve uses the OPTICS clustering algorithm~\cite{Ankerst1999OPTICS} \rev{to generate clusters to color the states by.}\redout{~color-codes each state based on its assigned cluster, and updates all of the visualizations with the new colors.}
Clustering states together based on their\redout{~analyst-defined} properties provides a slow, but\redout{~fairly accurate and} flexible method to directly compare trajectories. 

\rev{These interactions were designed to replace one of the inherent visual features of chord diagrams in an earlier design, where regions could be rendered as arcs on a circle and linked together based on similarity. 
When we attempted to implement chord diagrams in MolSieve, we found them to be cluttered and confusing when trying to interpret the temporal structure of the data.}

\rev{Clicking anywhere in the Super-State View updates the State Detail Widget with the state that occurs most frequently within the region, referred to as the region's \emph{characteristic state}. 
Characteristic states describe the general properties of these regions ~\cite{shalloway1996macrostates}. 
Thus, this interaction allows analysts to quickly determine if any structural changes occurred between super-states.
Once a change has been identified, analyst can seek more detailed information about the change within the Transition Region View between the two differing super-states.}
\definecolor{yellow}{HTML}{FFC40C}
\section{Case Studies}
We demonstrate the efficacy of MolSieve by presenting two case studies in which we conducted pairwise analysis~\cite{arias-hernandez2011pairanalytics} with our domain experts E1 and E2.
The first case study involves analyzing two long-duration trajectories of platinum nano-particles, first by determining sub-regions in each trajectory where the particle undergoes a structural change and then comparing the\rev{m}.\redout{sub-regions from each trajectory together.}
The second case study focuses on atom vacancy analysis, where a reference atomic configuration is compared to states within the trajectory. 
In atom vacancy analyses, experts typically look for regions within a simulation where the ``missing" atoms begin to displace in tandem. 

\subsection{Platinum Nano-particles}\label{subsec:platinum_nano_particles}
Our nano-particle expert (E1) aimed to identify and characterize significant fluctuations in the shape of a platinum nano-particle subjected to high temperatures.
\redout{The analyst~}\rev{E1} began the case study by loading a simulation of a platinum nano-particle at 750 kelvins, which consists of approximately eighteen million transitions and twenty-five thousand unique states \rev{(Table~\ref{tab:speed_table})}, with each state representing different configurations of a nano-particle with 147 platinum atoms.

Based on a prior study of nano-particles~\cite{huang2018shapefluctuation}, the analyst decided that the best properties to analyze this simulation were the Common Neighbor Analysis (CNA)~\cite{honeycutt1987molecular}, Ackland-Jones~\cite{ackland2006applications} (AJ), and Polyhedral Template Matching~\cite{larsen2016robust} (PTM) atom characterization counts.
These analyses attempt to characterize the structure of a nano-particle based on descriptors of the local environment around each component atom and have been found to be strong indicators of transition regions.
The analyst wrote a script that used OVITO~\cite{stukowski2010atomistic} to compute these properties and loaded them into MolSieve (\textbf{R2}).
\rev{Since it was difficult to tell what was occurring to the nano-particle from the default 3D render, our analyst wrote a visualization script that highlights CNA counts within states (Figure \ref{teaser}.6a,~\ref{teaser}.6b,~\ref{teaser}.7,~\ref{teaser}.11a, ~\ref{teaser}.11b, ~\ref{teaser}.14). 
The CNA visualization script renders HCP atoms as \textcolor{red}{red}, ICO atoms as \textcolor{yellow}{yellow}, and FCC atoms as \textcolor{green}{green}.}

\subsubsectionheader{Identify Transition Regions (T1):}
E1 decided to load the trajectory with a GPCCA clustering range of 2-20 and a simplification threshold of 0.75. 
GPCCA\redout{~determined that the optimal clustering for this trajectory used 2 clusters} \rev{split the trajectory into two clusters}, yielding a small red cluster in between a dominant teal cluster (Figure~\ref{teaser}.1) which E1 zoomed in on using the Timeline View.
This revealed a busy region with many possible transitions; however, the Super-State Views showed that the super-state distributions did not vary greatly between each other, so the analyst increased the number of clusters to 4, hoping to reveal more fine-grained super-states (Figure~\ref{teaser}.2).
Once the simplification was rendered, they found that there were a number of transition regions between super states where the ICO and HCP counts of the nano-particle were rising (\textbf{R1}; \rev{(Figure~\ref{teaser}.3a and~\ref{teaser}.3b)}.\redout{~demonstrate Super-state Views that exemplify this change.}

\subsubsectionheader{Analyze Transition Patterns (T2):}
\redout{Since each analysis produces different results for each structural type,~}\rev{T}he analyst added a multi-variate control chart using the ICO counts from all three analyses to see if they would all point towards the same regions (Figure~\ref{teaser}.4a and~\ref{teaser}.4b). 
The analyst then found two sub-regions within a Transition Region View where the control charts indicated that the structure of the nano-particle\redout{~suddenly} changed (Figures~\ref{teaser}.5a,~\ref{teaser}.5b; \textbf{R3}).

Next,\redout{~the expert decided to click} \rev{E1 clicked} on the Super-state Views (Figure~\ref{teaser}.3a,~\ref{teaser}.3b) surrounding that Transition Region View to get an understanding of how the nano-particle changed from the first super-state to the second; the characteristic states of each super-state are shown in Figures~\ref{teaser}.6a and~\ref{teaser}\rev{.}6b. 
\rev{Since it was difficult to tell what was occurring to the nano-particle from the default 3D render, the analyst changed the 3D view to highlight CNA counts.
This revealed a sudden change in the \textcolor{green}{ICO} count, where the two green atoms disappear in Figure~\ref{teaser}.3a disappear.
}
\redout{this view is included alongside the default 3D render throughout the Figure (1.6a, 1.6b, 1.7, 1.11a, 1.11b, 1.14). 
The CNA visualization script renders HCP atoms as red, ICO atoms as yellow, and FCC atoms as green.}

To verify that the sudden change in ICO count was not a random event, they double-clicked the Transition Region View to expand it.\redout{~on both sides.} 
This confirmed that the nano-particle stays in the same configuration for some time before suddenly undergoing a drastic change in the FCC and ICO counts.
Satisfied, they made a selection in the region where the ICO count suddenly changed from zero to one (Figure~\ref{teaser}.5a), which rendered a Sub-Sequence component. 
Then, they clicked through the states in the Sub-Sequence Component to get a detailed look at what was occurring to the particle (\textbf{R4}).
This revealed that the trajectory was undergoing a transformation (Figure \ref{teaser}.7) within the region the analyst selected (Figure~\ref{teaser}.5a); the nano-particle started the transition with two FCC atoms (Figure~\ref{teaser}.6a) and lost them (Figure~\ref{teaser}.6b).
They ignored the other sub-region where the ICO count dropped (Figure~\ref{teaser}.5b), stating that ``This is normal behavior in simulations with a heterogeneous energy barrier: the system tries to escape its configuration but is not able to, causing it to change before returning to its previous configuration; this is why I wanted to check the region on the left."

\begin{figure*}[t]
    \centering
    \includegraphics[width=\textwidth]{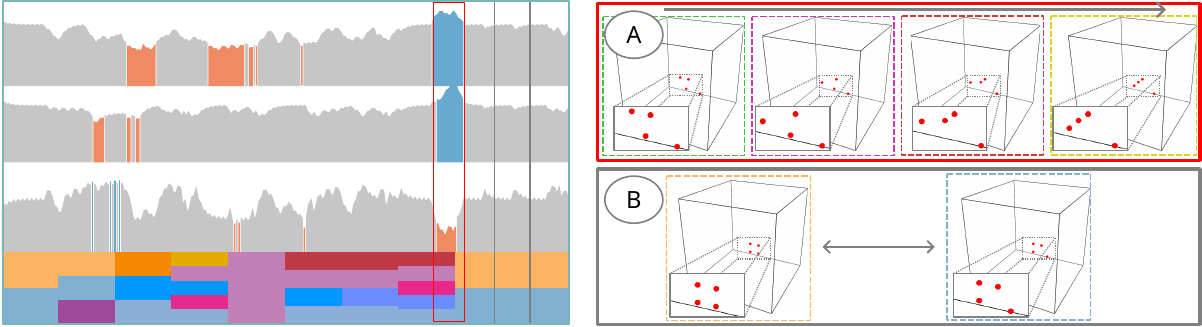}
    \vspace{-4mm}
    \caption{Results of the defect analysis case study. (A) Demonstrates an example of the diffusive transitions discovered within the simulation, which are a set of unique structural changes occurring to the defective region within the Tungsten crystalline lattice. (B) Demonstrates an example of "fluttering", where the defects within the lattice move back and forth between two configurations, one atom at a time. These kinds of transitions are the predominant transformations occurring to the lattice throughout the trajectory.
    The dashed rectangles represent the colors of each state's ID, and demonstrate that the State Space Charts were effective in capturing the regions of interest.}
    \vspace{-4mm}
    \label{case_study2}
\end{figure*}
 
Once the transition was found, they decided to run Nudged Elastic Band (NEB) calculations on the both ends of the suspected transition region (\textbf{R5}).
The NEBs confirmed that the transition to and from the suspected transition region took a large amount of energy, thus\redout{~confirming the transition region's accuracy and} demonstrating that our system is a significant improvement in terms of detecting regions of interest in large molecular dynamics simulations.

\subsubsectionheader{Ensemble Analysis (T3):}
Once the transition was confirmed, the analyst decided to load another platinum nano-particle trajectory at 800K. 
E1aimed to determine if the structural changes they observed in the particle at 750K were similar to the ones observed at 800. 
The 800K simulation contains thirteen million transitions and fifty-three thousand unique states \rev{(Table~\ref{tab:speed_table})}.

Once the simulation was loaded, the analyst used a similar workflow to determine where the transition regions occurred in the trajectory by carefully adjusting the simplification threshold until a suitable number of possible transition regions were displayed. 
Starting at the simplification threshold's default value of 0.75 did not yield any transition regions; however, it led the analyst to zoom into a sequence of super-states where the ICO count was changing from zero to one.
Increasing the simplification threshold to 0.85\redout{~showed that the super-states contained regions between them that reflected a similar type of behavior} \rev{revealed super-states undergoing a transition similar} to the 750K trajectory (Figure~\ref{teaser}.8).
The state IDs\redout{~heavily} overlapped between the two trajectories, and\redout{~there were a number of} \rev{many} regions that contained the same 4 states that the 750K simulation spent large amounts of time in \redout{as seen in the State Space charts} \rev{(Figure~\ref{teaser}.12a and~\ref{teaser}.12b}; \textbf{R1}, \textbf{R3}). 

The analyst then decided to click the \inlinebutton{figures/buttons/find_region_withtext.pdf} button and select the transition region they discovered in the 750K trajectory.
This revealed many regions shared a large portion of states with the selection,\redout{~so the analyst singled out the highest scoring regions, one of} \rev{a region} which scored 12\% similarity based on the set of unique states present in each region, which can be seen on the Timeline View (Figure~\ref{teaser}.9) (\textbf{R4}).
\rev{A similarity of this magnitude is significant due to the fact that simulations are unlikely to contain the same states in a small temporal region.}
They then used the \inlinebutton{figures/buttons/compare_region_withtext.pdf} button to examine the difference in distributions between the regions that scored highest on similarity and the original transition region they discovered (Figure~\ref{teaser}.10).
While the region that was 12\% similar did not have the same transition characteristics, the analyst found a region that had a similar shift in its ICO and HCP count.
Moreover, when the analyst clicked on the two Super-State Views surrounding the region, they found that the first super-state had the same characteristic state as the first in the previously found region, and the second super-state was a rotation of the previous tailing super-state (Figure~\ref{teaser}.12a, \ref{teaser}.12b).

While the nature of the transition was similar based on the control charts, E1 wondered if the states were truly structurally similar, so they went to use the \inlinebutton{figures/buttons/state_clustering_withtext.pdf} button. 
Recoloring the states based on their structural cluster revealed that the region shared\redout{~a number of structurally similar} \rev{many} states, particularly around the sub-regions where the analyst believed a transition was occurring (Figures~\ref{teaser}.12a and \ref{teaser}.12b). 
The analyst also used the Sub-Sequence Comparison Widget to compare the two selections (Figure~\ref{teaser}.13), which verified that the transitions were similar in nature as the states were rotational analogs of each other.
The combination of these comparisons reassured the analyst in their conclusion that these transitions were of a similar nature (\textbf{R4}). 
While not identical to the one found in the 750K simulation, the sub-region found by the analyst also describes how the nano-particle loses FCC atoms and gains an ICO atom (Figure~\ref{teaser}.1\rev{4}\redout{5}); this slight difference is to be expected due to the fact that MD simulations are stochastic by nature.
This discovery demonstrates that our system is effective in not only detecting regions of interest in one long-duration simulation but is also capable of\redout{~comparing and describing} \rev{detecting} similar physical occurrences in multiple simulations.

\subsection{Bulk Tungsten Defect Analysis}\label{subsec:defectanalysis}
The goal of a defect analysis is to understand the way the point defects in a crystalline structure evolve over the course of a simulation; these defects determine the properties of a given material.
Typically, analysts use the Wigner-Seitz cell method~\cite{zou1994topological} to visualize the difference between a state in a defect simulation and a reference structure that does not have any defects. 
Our analyst, who specializes in cell defects (E2), provided a reference Tungsten lattice with 2,000 atoms, which represented a perfect, defect-free crystalline structure, as well as a Python script from his daily workflow that compares a state and the reference structure using the Wigner-Seitz analysis.
The script they provided outputs the defective atoms in each state and displays them, which the analyst used as the state view for the case study, seen in the renders for Figures~\ref{case_study2}.A and~\ref{case_study2}.B.
Additionally, the analyst used the output from the script to create three properties which described the center of mass of the atoms that were defective (\textbf{R6}). 

To begin the case study, the analyst loaded their scripts and a simulation of a Tungsten crystalline lattice being subject to various deformations at 1000 kelvin. 
This data-set was considerably smaller than the nano-particle case study, having only approximately 800 transitions and only 50 unique states \rev{(Table~\ref{tab:speed_table})}. 
However, the size of each state was considerably larger, as each state represented a Tungsten lattice with 1996 atoms.

\subsubsectionheader{Analyze Transition Patterns (T2):}
MolSieve initially classified the entire trajectory as 3 super-states, which meant that the GPCCA simplification was not useful for this data-set. 
This prompted\redout{~the analyst} \rev{E2} to set the simplification threshold to 1.0,\redout{~effectively turning off the simplification feature} and rendering all of the GPCCA clusters as transition regions, allowing the analyst to see the control charts for each property.
Once\redout{~the trajectory} \rev{it} was re-rendered, the analyst noticed that the moving average time period for each Transition Region View was very high, obscuring potentially interesting transitions\redout{~(i.e., over-smoothed to the extent that no interesting regions pop out)}, so they set the moving average time period for each transition region to 10.
Once the system was configured properly,\redout{~the transitions of interest within the trajectory became apparent, as} the control charts exposed regions where the center of mass changed rapidly in all three dimensions.
MolSieve immediately identified diffusive transitions (Figure~\ref{case_study2}.A), highlighting them among the numerous repetitive thermal vibrational motions (Figure ~\ref{case_study2}.B) that were composed of single vacancies moving back and forth (\textbf{R1}, \textbf{R3}). 
E2 then selected several regions highlighted by the control charts and was able to identify and follow the chain of events for several diffuse transitions. 
This case study was able to demonstrate that MolSieve is effective in finding regions of interest in diverse analysis scenarios. 

\subsection{Domain Expert Feedback}
To evaluate MolSieve, we conducted an \rev{hour-long} semi-structured interview session\redout{~(one hour long)} with E1 and E2.\redout{~the two domain experts who had conducted our case studies (E1, E2).}
During the interview, we asked them to compare their daily workflow to using MolSieve and \rev{solicited their suggestions on improving the system.}\redout{~how it can be improved to better adapt to the problems they are currently facing.}

A typical workflow for a molecular dynamics analyst consists of running scripts for several days on simulation data and sifting through states manually\redout{~interesting data is found}. 
They typically visualize the states in OVITO\textcolor{red}{~\cite{stukowski2010atomistic}} and then click frame-by-frame to get an idea of what changes the system is going through.
The greatest challenge in analyzing simulations this way is the amount of data that needs to be processed which makes it difficult to keep track of transitions and one's temporal context within the trajectory.
E1, our nano-particle expert, remarked that ``The overall layout of MolSieve makes it easy to analyze these data-sets. It is very easy to understand where you are in the trajectory, just by looking at the Timeline View. This helps me think about what is going on in the simulation as a whole, and I don't feel like I have tunnel vision while examining data."
They continued their reflection on the system by comparing the experience of examining regions of interest in MolSieve with their daily workflow, specifically praising the 3D overview within Sub-Sequence Components, saying that ``The 3D overview [within the Sub-Sequence Component] provides a very nice, pictorial, visual effect that gives a preview of what the particle is going through. I don't have to waste time clicking back and forth between states to get an idea of what I'm looking at."
E2, our defect analysis expert, reflected on MolSieve's visual design by saying, ``The combination of the control charts and the aggregate state space chart make it easy to find regions of interest within a transition region. The aggregate state chart also tells me which regions to avoid selecting, since it's so easy to see where the simulation gets stuck jumping between a small set of states."

The experts found that MolSieve was\redout{~extraordinarily} efficient, providing a massive productivity increase over their accustomed workflows.
E1 said, ``The system is exciting, as it takes an unimaginable amount of data and makes it interpretable. The nano-particle simulations we examined could take several lifetimes to sift through, and MolSieve manages to make it look trivial, with near real-time performance."
E2 added, ``The amount of data I was able to comb through with MolSieve would have normally taken a few weeks to do, and I managed to do this in just a few minutes," which indicates that we fulfilled \textbf{R7}.

The customizability of the system was a major selling point, as E2 stated, ``That is what really makes it come to life - this makes it applicable for a wide array of applications and will save us a considerable amount of time in the future." 
E2 continued the discussion by suggesting that analysts should be able to customize the simplification algorithm. 
This idea stems from the results of the atom vacancy case study (Section~\ref{subsec:defectanalysis}), \rev{where the simplification algorithm failed to produce transition regions.}
\redout{In the case study, the states were all strongly clustered together, which caused the simplification algorithm to fail to produce transition regions.}
To get around this, E2\redout{~simply} increased the simplification threshold to include the entire trajectory.
E2 warned that, "In principle, the simplification scheme in MolSieve should work on most data-sets but molecular dynamics simulations are often analyzed in various modalities, some of which are not captured by dividing the trajectory using GPCCA."
Thus, allowing experts to customize how the trajectory is simplified could make it easier to find relevant regions for various analyses.\redout{, as well as reduce the computational costs of analyzing large simulations that do not cluster well with GPCCA.} 
E2 added that the distance metric used in both the overview in the Sub-Sequence Component (Section~\ref{subsec:sub-sequence-panel}) as well as the heatmap in the Sub-Sequence Comparison Widget did not effectively describe the difference between two states.
This was due to the fact that we were studying the \textbf{absence} of atoms within a state.
To make these comparisons more useful, they suggested that the distance functions in MolSieve should also be customizable.\redout{~and provided an example distance function that used the centers of mass from the vacancies to compare states.}

Finally, E1 felt that the MolSieve was lacking a feature for comparing multiple individual states. 
We focused on comparing sub-regions within trajectories and did not consider the importance of being able to easily compare two or more states.\redout{~next to each other.} 
The Sub-Sequence Comparison Widget supports this to a limited extent, but E1 suggested an interaction that could ``save" states and show them \rev{on demand}.\redout{~when the side-bar is summoned.}
\redout{This would further reduce the cognitive load in interpreting regions of interest within trajectories and is a priority for future work.}
\section{Conclusion}
\begin{table}[]
\centering
\resizebox{\columnwidth}{!}{%
\begin{tabu}{crrrrrr}
\textbf{Simulation}  & \textbf{\begin{tabular}[c]{@{}c@{}}Generation \\ time (s)\end{tabular}} & \textbf{\begin{tabular}[c]{@{}c@{}}Simulation \\ time (ns)\end{tabular}} & \textbf{\# Timesteps} & \textbf{\# States} & \textbf{\begin{tabular}[c]{@{}c@{}}Time to load \\ (cached) (s)\end{tabular}} & \textbf{\begin{tabular}[c]{@{}c@{}}Total \\ preprocessing \\ time (s)\end{tabular}} \\
\midrule
\textit{nano-pt-700} & 35,994                                                                  & 62,857.99                                                                & 6,711,821             & 16,631             & 5.876                                                                            & 459.156                                                                             \\
\textit{nano-pt-750} & 35,992                                                                  & 49,869.45                                                                & 18,463,872            & 24,457             & 7.714                                                                           & 1507.284                                                                            \\
\textit{nano-pt-800} & 35,993                                                                  & 43,152.76                                                                & 13,348,978            & 53,018             & 10.489                                                                           & 1150.753                                                                            \\
\textit{nano-pt-900} & 57,586                                                                  & 31,636.00                                                                   & 7,721,529             & 490,226            & 14.979                                                                           & 9172.017                                                                            \\
\textit{tungsten}    & 10,800                                                                  & 10,000.00                                                                   & 866                   & 241                & 2.000                                                                             & 6.800                                                                              
\end{tabu}%
}
\caption{\rev{Several simulations that were tested on MolSieve are presented here. This table displays the total time it took to generate each simulation in ParSplice, the length of time the simulation represents in nanoseconds, the number of discrete timesteps in the simulation, the number of unique states, the time it takes to load the simulation when cached, and how long it takes to load each simulation.}}
\vspace{-4mm}
\label{tab:speed_table}
\end{table}

In this work, we present MolSieve, a visual analytics system for \redout{analyzing} long-duration molecular dynamics simulations modeled by discrete Markov chains. 
Through the use of multiple coordinated visualizations powered by a data simplification scheme unique to MD simulations, MolSieve makes it possible to analyze previously unexplored simulation data-sets\redout{of unlimited size}. 
The comparison interactions offered by the system provide support for analyzing simulation ensembles.
Additionally, MolSieve's Python programming interface lets it accommodate a wide variety of simulations. 
To demonstrate the effectiveness of MolSieve's design, we analyzed three simulations alongside our domain experts: two nano-particle simulations and one atom vacancy simulation.
\rev{Table~\ref{tab:speed_table} provides a detailed look at the efficiency of the system}.
\redout{We found that MolSieve was capable of efficiently extracting insight from both types of simulations.}

However, it became apparent that some of its components need to support further customization.  
We found that the simplification algorithm would sometimes return many regions in a trajectory, which led to the screen being highly cluttered.
This would require the analyst to zoom in using the Timeline View to get a better idea of the general trend within the trajectory. 
This can be mitigated by reformulating the way regions are rendered to only show large regions until the zoom level is appropriate.
We also found that some visual design elements\redout{~need to} \rev{must} be adjusted\redout{~to better serve experts}; these issues are particularly prevalent in the color encodings of the interface.
The analysts found that coloring states by their IDs made it difficult to distinguish them from one another once a large number of states were rendered on the screen, which we attempted to remedy by implementing the state clustering function. 
However, the state clustering function would sometimes also have color overlap, which could be reduced by mapping the number of clusters to a set of salient colors. 
Alternatively, we could explore using different visual encodings to distinguish a large number of classes.
\rev{Another limitation is the inability to view a list of the most frequently occurring states within a Super-State.
This can be addressed by adding a widget that shows all of the most frequently occurring states in a region.
}

In the future, we plan to address some of the limitations of the system, including the cramped visual encoding space and the need for extra customization.
Providing additional support for exploring biological simulations would be of particular interest, as this could lead to a truly general MD region-of-interest visual analytics system.
To continue scaling, we plan to switch the rendering engine to use WebGL instead of SVG, allowing MolSieve to take advantage of the current innovations in consumer graphics technology. 
\rev{Moreover, a number of techniques have yet to be integrated into our system - improving the 3D rendering pipeline will allow MolSieve to support a number of novel analyses (e.g.,~\cite{Wu.2022.VAD, Tian.2023.LVA, Kern.2023.AVI} and rendering techniques~\cite{Schatz.2019.IVB}.}
Future work will also include a method to recall expert selections, a direct state comparison view, and better 3D rendering support.

\section{Supplementary Materials}
\label{sec:supplemental_materials}
We included a demo video that showcases the first case study and an instruction manual for MolSieve as supplementary material.
MolSieve's source code is available at \url{https://github.com/rostyhn/MolSieve}.
\acknowledgments{
This project was supported by the Exascale Computing Project (17-SC-20-SC), a collaborative effort of the US Department of Energy Office of Science and the National Nuclear Security Administration and the U.S. Department of Homeland Security under Grant Award Number 17STQAC00001-06-04. Los Alamos National Laboratory is operated by Triad National Security LLC, for the National Nuclear Security administration of the U.S. DOE under Contract No. 89233218CNA0000001. We graciously acknowledge computing resources from the Los Alamos National Laboratory. The views and conclusions contained in this document are those of the authors and should not be interpreted as necessarily representing the official policies, either expressed or implied, of the U.S. Department of Homeland Security.
We also would like to acknowledge Jiayi Hong and Andrew Garmon for their discussions and contributions to the paper.
}

\bibliographystyle{abbrv-doi-hyperref}

\bibliography{bibliography}
\end{document}